\def\agt{\,\raise.3ex\hbox{$>$\kern-.75em\lower1ex\hbox{$\sim$}}\,}
\def\alt{\,\raise.3ex\hbox{$<$\kern-.75em\lower1ex\hbox{$\sim$}}\,}

\def\har{\hat{r}}
\def\hra{\hat{r}_{\rm a}}
\def\hrp{\hat{r}_{\rm p}}
\def\nuk{\nu_{\rm K}}
\def\nukN{\nu_{\rm K,N}}
\def\rp{r_{\rm p}}
\def\Rp{r_{\rm p}}
\def\Ra{r_{\rm a}}
\def\ra{r_{\rm a}}
\def\r1{r_{\rm a}}
\def\brp{\bar{r}_{\rm p}}
\def\bra{\bar{r}_{\rm a}}
\def\br{\bar{r}}
\def\Re{R_{\rm e}}
\def\risco{r_{\rm isco}}

\def\nupa{\nu_{_{\rm PA}}}
\def\nus{\nu_{_{\rm s}}}

\def\nunp{\nu_{_{\rm NP}}}

\def\nuisco{\nu_{\rm isco}}

\documentstyle[epsfig]{mnusa}

\title{Bound near-equatorial orbits around neutron stars}
\author[D. Markovi\'c]{D. Markovi\'c \thanks{Also Department of Physics.} \\
    Center for Theoretical Astrophysics, University of Illinois at Urbana-Champaign,
    1110 West Green Street, Urbana, IL 61801, USA\\
        draza@qpo.physics.uiuc.edu}
\begin{document}
\input{epsf}

\maketitle

\begin{abstract}
Recent discovery of kilohertz quasi-periodic brightness oscillations of
low mass X-ray binaries (LMXBs) has attracted attention to 
highly relativistic periodic motion near accreting neutron stars.
Most models proposed so far involve (almost) inertial motion in the
vicinity of the stars' innermost stable circular orbits.
In the present paper we study general-relativistic
circular and eccentric orbits around spinning neutron stars assuming
the orbits are slightly tilted with respect to the stars' equatorial planes.
We develop  analytical and numerical techniques for integrating bound 
timelike geodesics in
fully relativistic neutron star
spacetimes obtained by modern numerical codes. 
We use equations of state of neutron star matter that span 
a broad range of stiffness, while the explored range of masses
($M > 1.7 M_{\odot}$) and spin frequencies ($\nus < 600\,$Hz) 
is motivated by the observations of LMXBs.
We investigate the general-relativistic effects of periastron advance
and nodal precession in the strong gravitational 
fields of rotating neutron stars  
and compare quantitatively the associated orbital frequencies with the 
more readily obtainable frequencies
of orbits around Kerr black holes on the one hand, 
and low-order post-Newtonian (PN)
expansions, on the other.  While Kerr results approximate the periastron
advance frequency much better than the PN expressions, the retrograde
torque caused by the high quadrupole moments of the rotating stars
clearly favours the PN approximation in the case of
nodal precession.  The methods developed in the present paper
are used in the companion paper to test recent hypotheses
regarding the origin of quasi-periodic oscillations in accreting
neutron star sources.

\end{abstract}


\section{Introduction and overview}

Over the past four years,  observations by the Rossi XTE satellite of
low-mass X-ray binaries (LMXBs) have produced an abundance of data on the 
rapid variations of these sources' X-ray brightness (see van der Klis 1998, 2000,
for an overview, and references therein).   Most prominent in the sources'
high-resolution power spectra, the kilohertz-range quasi-periodic oscillations
(QPOs) have been detected in $\sim20$ binaries with compact accretors identified
as neutron stars.  The kHz QPOs often occur in pairs at frequencies $\nu_1$ and
$\nu_2 > \nu_1$ (with $\nu_2$ ranging between $500\,$Hz and $1200\,$Hz)
that can vary in a single source by up to a factor of 
$\sim2$ (over as little as few minutes)
in step with an
inferred change in the accretion rate.  In addition to the kHz QPOs,
the horizontal-branch oscillations (HBOs) of the high-luminosity (near Eddington)
neutron-star Z sources and
`bumps' in the power spectra of the lower-luminosity
(between $10^{-3} L_{\rm Edd}$ and a few $10^{-1} L_{\rm Edd}$)
atoll sources 
(for the definition and properties of the Z and
atoll sources see Hasinger \& van der Klis 1989)
have been observed at lower frequencies
$10\,{\rm Hz} \alt \nu_{\rm L} \alt 60\,$Hz.
 These frequencies also vary in apparently tight correlation
with the kHz QPO frequencies.

The fact that the kHz QPO frequencies lie in the range of orbital frequencies of
bound geodesics near neutron-star-mass objects [the Kepler frequency
for the innermost stable circular orbit (ISCO) around a static, spherical
star of mass $M$ is $\nu_{\rm isco} = 1100\,{\rm Hz}\, (2 M_{\odot}/M)$]
motivated several theoretical scenarios (see, e.g., 
Strohmayer {\it et al.} 1996, Miller, Lamb
\& Psaltis 1998, Stella \& Vietri 1999) for the X-ray luminosity modulation
that all involve inertial or near-inertial 
motion of localised condensations --- variously named `clumps'
or `blobs' --- in the accretion flows around neutron stars.  For instance,
in a purely inertial (geodesic) 
orbital model,
Stella \& Vietri (1999; see also Karas 1999) have proposed for a given source
an identification of the variable QPO frequencies
$\nu_2$ and $\nu_1$
 with, respectively,
the Kepler frequencies $\nuk$ and the periastron advance frequencies $\nupa$
of a family of eccentric orbits.
The preferred plane of accretion (and, thus, of the clumps' motion) is
normally assumed to lie close to the
rotation equator of the central neutron star,
a natural consequence of accretion-driven neutron star spin-up (similar reasoning
clearly favours {\it prograde} motion in the accretion flow).  However, we
need not exclude {\it small} deviations of the orbits from the equatorial 
plane, and so it has been speculated that a signal ---
e.g., the QPOs observed at frequency $\nu_{\rm L}$ --- might be produced by the
nodal precession of the orbital planes at frequency $\nunp$ or
its first overtone (see, e.g., 
Stella \& Vietri 1998 and 
Morsink \& Stella 1999 for the nodal precession of circular orbits).

If any of the orbital models is confirmed, 
observation of orbital frequencies would provide
a direct probe of the highly curved spacetimes
around rapidly spinning neutron stars and an 
unprecedented test of the general theory
of relativity in the strong-field regime.
The highly accurate measurements of the kHz QPO frequencies (e.g., the relative error
$\alt1\%$ for Sco X-1; van der Klis {\it et al.} 1997) impose, however, stringent requirements
on the QPO models; they also make it necessary to know the theoretical 
orbital frequencies with an accuracy that may render post-Newtonian 
approximations inadequate.  Similarly, the spacetimes around neutron stars
can differ from the Kerr black-hole  spacetimes to such a degree as to
invalidate the use of Kerr-based orbital frequencies. 

It is our purpose in the present and a companion 
paper to investigate in some detail orbital motion
near the equatorial planes of neutron stars and to test carefully the agreement between
the frequencies of bound orbits with the  QPO frequencies observed so far in neutron-star
LMXBs. Specific QPO models will be compared with the existing
data in the companion paper (Markovi\'c \& Lamb 2000).   In the present paper we
study the motion of test particles on orbits of
arbitrary eccentricity around rapidly spinning neutron stars.
We pay particular attention to the general-relativistic effects of periastron advance
and nodal precession in the neutron stars'
strong gravitational fields, 
and compare quantitatively the orbital frequencies associated with these effects with the
more readily obtainable frequencies
of orbits around Kerr black holes on the one hand,
and low-order post-Newtonian (PN)
expansions, on the other. 
As the basis of our study, we
set out to develop analytical and numerical techniques for
a sufficiently accurate computation of the frequencies
of orbits in fully general
relativistic numerical neutron star spacetimes.   
Modern numerical codes (Komatsu, Eriguchi \& Hachisu 1989a,b; 
Cook, Shapiro \& Teukolsky 1992, 1994a,b; Stergioulas \& Friedman 1995;
Nozawa et al. 1998)
allow an accurate computation of equilibrium uniformly spinning neutron star models
starting from an arbitrary tabulated equation of state (EOS), and it is
one variant (Stergioulas \& Friedman 1995) that we use to obtain external 
neutron star spacetimes in which we then integrate the equations of motion
for bound geodesics.  


The numerical neutron star models discussed in this paper are
all based on three equations of state that sample a broad range
of stiffness,  from the moderately
stiff EOS C (Bethe \& Johnson 1974;  see also the early review by
Arnett \& Bowers, 1977) 
to the 
very stiff EOS L \cite{L}.  The classic EOS C describes cold nucleon ({\it n} and {\it p})
+ hyperon ($\lambda$, $\Sigma^{\pm, 0}$, $\Delta^{\pm, 0}$ and $\Delta^{++}$)
matter with the nucleon-nucleon potential represented by a set of Yukawa functions.
The high-density behaviour is dominated by the strong short-range repulsion due
to the $\omega$-meson exchange.   EOS C allows the maximum mass of $1.86 M_{\odot}$
for non-rotating stars.   On the other hand,  EOS L, based on coupling
of (non-relativistic) nuclear matter to a mean scalar field, 
provides for a much higher maximum mass, $2.71 M_{\odot}$, for non-rotating neutron stars and,
accordingly, larger equatorial radii.  For the third EOS of intermediate stiffness
we choose EOS UU \cite{UU}, which, besides including the nucleon-nucleon potential
(specifically, the Urbana $v_{14}$), achieves the relatively
greater stiffness at high density due to the Urbana VII 
three-nucleon potential
(Schiavilla, Pandharipande, \& Wiringa 1986).  
The maximum non-rotating NS mass
for this EOS is $2.20 M_{\odot}$.    
[The more recent EOS A18+UIX + $\delta v_b$ 
of Akmal, Pandharipande and Ravenhall (1998; see also 
Pandharipande, Akmal \& Ravenhall, 1998) is based on the modern Argonne $v_{18}$
two-nucleon potential and the Urbana IX three-nucleon potential and allows for the
dependence of the interaction between two nucleons
on their {\it total} centre-of-mass
momentum; its stiffness is similar to that of EOS UU and the maximum non-rotating
stellar mass is $2.23 M_{\odot}$.]
These three equations of state span the entire range of EOS stiffness that can provide
the neutron star masses ($1.7 < M < 2.2$; from now on all masses will be given in units
of solar mass)
and radii required by the QPO models discussed in
the companion paper ~\cite{M1999}.


In Section 2, we derive the equations of motion for a general bound
geodesic near the equatorial plane of the stationary
spacetime around a neutron star.  The equations
of motion involve the values of four radius-dependent
potentials (provided as a tabulated output from a NS-model numerical code) in the
equatorial plane, as well as the potentials' second derivatives with respect
to the vertical coordinate.
The presence of two integrals of motion, the energy $E$ and the vertical
component, $L$, of the  angular momentum,
allows integration of the radial and azimuthal 
equations up to quadratures (we specifically solve for {\it prograde}
orbits).  For orbits (slightly) tilted
relative to the equatorial plane, the lack of a third integral of motion leaves one
with two first-order
Hamilton's equations for the vertical $\theta-$motion.
The precession of the orbital plane is then most conveniently analysed
by means  of Floquet's method.  In the balance of Section 2 we discuss
similar motion in the Kerr black hole spacetime, where the presence
of a third integral of motion, the Carter constant, simplifies the
treatment considerably.   Finally, we briefly discuss low-order
post-Newtonian expressions for the periastron advance 
and nodal precession frequencies.

In Section 3, we illustrate, by means of several examples,
the salient features of the orbits around neutron stars, devoting
most of our discussion to the dependence of the orbital frequencies, $\nuk$, $\nu_r =
\nuk - \nupa$ and $\nunp$, 
on the radii of circular orbits or the
periastra and apastra of eccentric orbits (unless specified otherwise,
all radii will be given in units $GM/c^2$).   We pay particular attention
to the manner in which the frequency profiles are affected by
equations of state of various stiffness.  For instance, the stiffer EOS yield 
neutron stars of greater extent,
which, in turn, can exclude highly eccentric orbits with
small periastra and low $\nu_r$.   The larger stellar extent (and, accordingly,
greater magnitude of the quadrupole moment $q G^2 M^3/c^4$) 
also reduces significantly the  nodal precession
frequencies for stars of given  mass $M$ and angular momentum $j GM^2/c$
(this effect was previously explored for circular orbits by 
Morsink \& Stella 1999).
We also assess the accuracy of approximating
the orbital frequencies in neutron-star spacetimes  
by the more readily computed
frequencies of orbits in Kerr spacetimes of the same mass $M$
and angular momentum $j G M^2 /c$,  or by the frequencies 
obtained from low-order
post-Newtonian expressions (including the relatively large mass quadrupole terms
characteristic of neutron stars).   The results of this paper are 
summarised and discussed in Section 4.

Equipped with accurately computed  frequencies of orbits
in neutron star spacetimes, we
proceed in the companion paper  \cite{M1999}
to investigate the possibility of their matching the
QPO frequencies of neutron-star LMXBs.

\section{Bound geodesics around spinning compact stars}

\subsection{Neutron stars}
The code of Stergioulas \cite{Sterg.95} yields the
metric around a neutron star in the form isotropic in the $\br-\theta$ sector
\begin{eqnarray}
\label{metric}
ds^2 &=& - e^{\rho + \gamma} dt^2 + e^{\gamma -\rho} \br^2 \sin^2 \theta 
        (d\phi - \omega dt )^2   \nonumber \\
     & &  + e^{2\alpha} (d\br^2 + \br^2 d\theta^2),
\end{eqnarray}
where $\gamma(\br,\theta)$, $\rho(\br,\theta)$, $\alpha(\br,\theta)$,
$\omega(\br,\theta)$ are the four potentials computed numerically.
The integration of orbital equations, described below, requires
the four potentials to be given as smooth functions of $\br$.
We achieve sufficient smoothness through a combination of spline
and high-order
Chebyshev polynomial representations (see, e.g., Press {\it et al.} 1992).
Throughout the paper, $r$ will denote only the coordinate-independent 
{\it circumferential} radius.  For metric (\ref{metric}),   
$r = \exp[ (\gamma -\rho)/2)] \br$.

\begin{figure*}
\label{orbits.ecc}
\centering
\hbox{\epsfxsize= 6 cm \epsfbox[45 320 225 615]{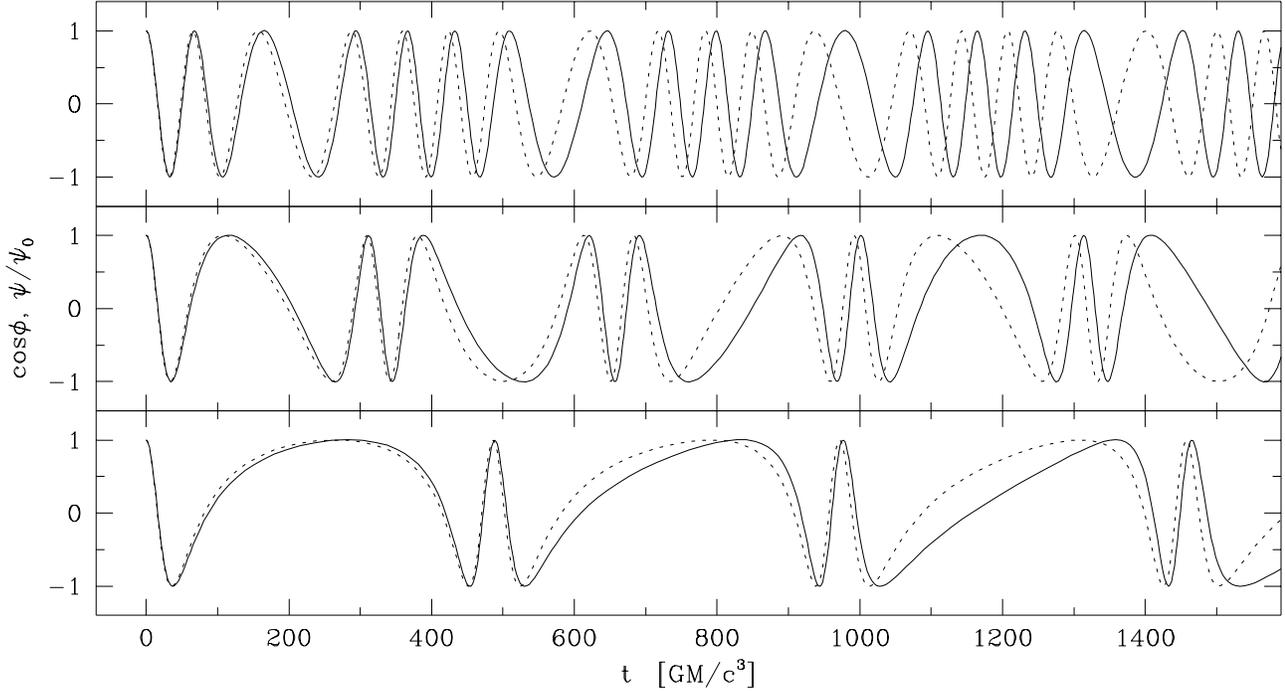}}
\caption{Plots of $\cos\phi$ (dotted lines) and $\psi/\psi_0$  (solid lines)
for three eccentric
orbits of common periastron $\rp = 4.7$ around
a neutron star of mass $M=1.8$, EOS UU, $\nus = 500\,$Hz ($j = 0.20$).
The apastra and the orbital frequencies are $\ra =$ 8.0
($\nuk = 1400\,$Hz, $\nunp = 38\,$Hz),
13.4 (821, 17)
and 23.0 (475, 8.6).
}
\end{figure*}

The metric components' independence of $t$ and $\phi$
implies that the components $u_t \equiv -E$ and $u_{\phi} \equiv L$
of the 4-velocity of a test particle 
are conserved.  It immediately follows that 
\begin{eqnarray}
\label{dtdphi}
\frac{dt}{d\tau} &=& e^{-(\rho + \gamma)} (E - \omega L)
	\nonumber \\
\frac{d\phi}{d\tau} &=&\frac{L}{\br^2 \sin^2 \theta} e^{\rho - \gamma}
        + \omega e^{-(\rho +\gamma)} (E - \omega L).
\end{eqnarray}

Assume for the moment a purely equatorial motion, $\theta = \pi/2$.
The normalisation $u^{\mu}u_{\mu}=-1$ 
leads then to the
equation for radial motion
\begin{eqnarray}
\label{drdtau}
\left(\frac{d\br}{d\tau}\right)^{2} &=& F(\br) 
          \nonumber \\
          &\equiv & e^{-2\alpha}\left[
         e^{-(\rho +\gamma)} (E - \omega L)^2 - \frac{L^2}{\br^2} e^{\rho - \gamma} 
	 - 1\right].
\end{eqnarray}

For given apastron radius $\bra$ and periastron radius $\brp$,
a bound orbit exists if one can find $L$ ($L>0$ 
for prograde orbits) such that $F_{\rm a} = F_{\rm p} =0$
(subscripts denote the values of $\br$ at which the quantities
are evaluated), i.e.,  such that solve the equations
\begin{equation}
\label{eq.L}
E = \omega_{\rm a} L + \left(A_{\rm a} L^2 + B_{\rm a}\right)^{1/2} =
      \omega_{\rm p} L + \left(A_{\rm p} L^2 + B_{\rm p}\right)^{1/2},
\end{equation}
where $A(\br) \equiv \exp[2\rho(\br)]/\br^2 $ and $B(\br) = 
\exp[\rho(\br) + \gamma(\br)]$.   
The required solution exists only if $\brp > \br_{\rm mb}$,
where $\br_{\rm mb}$ is the periastron of the marginally
bound ($E=1$) orbit.   Generally, for $L$ and $E$ corresponding
to a bound orbit, there also is a third root of the equation $F(\br_3) = 0$,
closer to the origin, $\br_3 < \brp$.  For the marginally
bound orbit $\brp = \br_3$, $\bra=\infty$  [$\,F^{\prime}(\brp)=0$,
 $F^{\prime\prime}(\brp) > 0\,$; the prime denotes the derivative
with respect to $\br$].

For a given $\brp < \br_{\rm isco}$, where $\br_{\rm isco}$ is
the radius of the innermost stable circular orbit, there is 
a minimum apastron, $\br_{{\rm a},{\rm min}}(\brp) > \br_{\rm isco}$
below which there exist no bound (prograde) orbits.  For $\brp > \br_{\rm isco}$
we naturally have $\br_{{\rm a},{\rm min}} = \brp$.  

Once $L$ and $E$ are found for given $\brp$ and $\bra$, it 
is straightforward to obtain the period (in the distant observer's time $t$)
of radial motion
\begin{equation}
\label{nur.NS}
\frac{1}{\nu_r} = 2\int_{\brp}^{\bra}  
           \frac{E -\omega L}{\sqrt{F}} e^{-(\rho +\gamma)} d\br,
\end{equation}
as    well as the frequency of azimuthal motion, i.e., 
the Kepler frequency
\begin{equation}
\label{nuk.NS}
\nuk = \frac{\Delta\phi}{\pi} \nu_r,
\end{equation}
where
\begin{eqnarray}
\label{dphi.NS}
\Delta\phi &=& \int_{\brp} ^{\bra} \frac{\dot{\phi}}{\dot{\br}}d\br
          \nonumber \\
          &=& \int_{\brp} ^{\bra}
              \left[ \frac{L}{\br^2} e^{\rho -\gamma} + 
        \omega e^{-(\rho +\gamma)} (E -\omega L)\right] \frac{d\br}{\sqrt{F}}.
\end{eqnarray}
\begin{figure}
\label{drift.ecc}
\centering
\hbox{\epsfxsize= 6 cm \epsfbox[50 200 410 680]{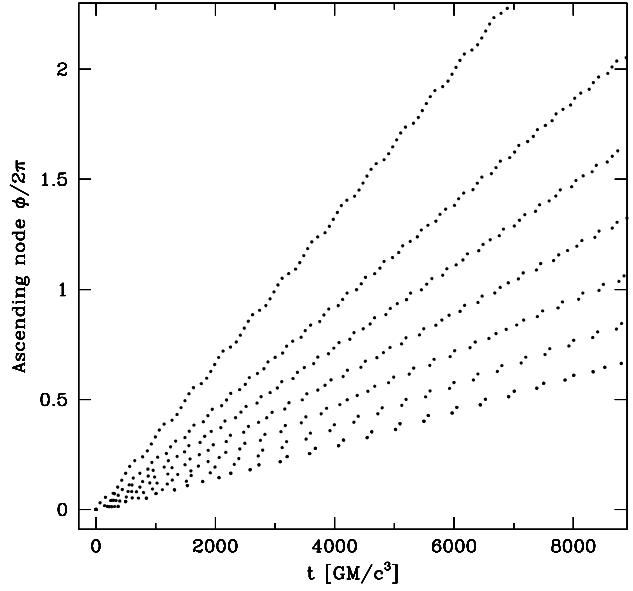}}
\caption{The drift of ascending nodal points for eccentric
orbits of common periastron $\rp = 4.7$ (the same NS as in Fig.~1; throughout
the paper $r$ denotes the {\it circumferential} radius)
and apastra  $\ra =$ 8.0
($\nuk = 1400\,$Hz, $\nunp = 38\,$Hz),  9.5 (1120, 26),
11.3 (964, 21),  13.4 (821, 17), 16.0 (692, 14), 19.2 (577, 11)
and 23.0 (475, 8.6).
}
\end{figure}
The periastron advance frequency is then simply
\begin{equation}
\label{nupa}
\nupa = \nuk - \nu_r.
\end{equation}


In the more general case of orbits out of the
equatorial plane, the motion is governed by the Hamiltonian
(see, e.g., Misner, Thorne \& Wheeler 1973)
\begin{eqnarray}
\label{H}
H &=& -\frac{1}{2} e^{-(\rho +\gamma)} (E - \omega L)^2 
       +\frac{1}{2}e^{\rho - \gamma}\frac{L^2}{\br^2 \sin^2 \theta}
      \nonumber \\
  & &     +\frac{1}{2}e^{-2\alpha} \left(p_{\br}^2 + \frac{1}{\br^2} p_{\theta}^2\right). 
\end{eqnarray}
For small displacements, $\psi \equiv \pi/2 - \theta$  ($p_{\psi} = -p_{\theta}$), 
from the equatorial plane, we can expand the four potentials, 
$\rho(\br,\psi) \approx \rho(\br,0) + \bar{\rho}(\br) \psi^2 /2 \equiv \rho_0 (\br)
 + \bar{\rho}(\br) \psi^2 /2 $, etc., 
and thus arrive at the `harmonic oscillator'
form quadratic in both $\psi$ and $p_{\psi}$
\begin{equation}
\label{H1}
H = H_0 (\br, p_{\br}) + \frac{1}{2}\Omega^2 (\br) \psi^2 + \frac{1}{2 I(\br)} p_{\psi}^2,
\end{equation}
where 
\begin{equation}
\label{I} 
I(\br) \equiv e^{2\alpha_0} \br^2
\end{equation}
and
\begin{eqnarray}
\label{Omega}
\Omega^2 (\br) &=&  e^{-(\rho_0 + \gamma_0)} \bigg[ (E - \omega_0 L)^2 
             \left( \frac{\bar{\rho}}{2} + \frac{\bar{\gamma}}{2} -
                   \bar{\alpha}\right) 
                \nonumber \\
         & &  \hspace{3cm} + (E - \omega_0 L) L \bar{\omega}\bigg]
               \nonumber \\
         & & \hspace{0.5cm}    +  e^{\rho_0 - \gamma_0} \frac{L^2}{\br^2} 
                 \left(1 + \frac{\bar{\rho}}{2} - \frac{\bar{\gamma}}{2} 
                  + \bar{\alpha}\right) + \bar{\alpha}.
\end{eqnarray}
Since the radial motion is (to the lowest order in $\psi$)  
given by equation~(\ref{drdtau}) [and thus $H_0 (\br, p_{\br}) = -1/2$], 
we have used this fact
to substitute $p_{\br} = \exp(2\alpha_0) d\br/d\tau$  in the expression
for $\Omega^2(\br)$.

The motion in the $\psi$ direction can now be computed from Hamilton's equations,
given here in the matrix form,
\begin{eqnarray}
\label{ham.eq}
& & \hspace{-0.4cm} \frac{d}{d\tau}\left(
 \begin{array}{c}
\psi \\
p_{\psi}
\end{array} \right)
= {\bf A} \left(
\begin{array}{c}
\psi  \\
p_{\psi}
\end{array} \right),
\nonumber \\  \nonumber \\
& & \hspace{-0.4cm} A_{12} = 1/I(\br), \hspace{0.4cm}  A_{21} =  -\Omega^2 (\br),
     \hspace{0.4cm} A_{11} = A_{22} = 0,
\end{eqnarray}
where $\br(\tau)$ is a solution of equation~(\ref{drdtau}).

The secular advance of the nodal line (i.e., the nodal precession)
is a straightforward consequence of the failure of a particle
to return to the initial $\psi$ after a full revolution
around the star.  For example, 
in Fig.~1 we plot $\cos\phi$ and $\psi/\psi_0$ for three orbits
of common periastron (circumferential) radius $\rp = 4.7$,
which
is smaller than $\risco = 5.4$ for the neutron star of mass
$M = 1.8$, EOS UU and spin frequency $\nus = 500\,$Hz ($j=0.20$). 
Notice the gradually increasing phase lag of the  
meridional motion relative to the azimuthal motion.  This leads
to a secular advance of the ascending node (see Fig.~2). 

For eccentric orbits, the rate of the advance varies somewhat over  
a few orbital periods $\nu_{\rm K}^{-1}$: since the radial frequency $\nu_r$ is
much smaller than $\nuk$ for orbits close to $\risco$ (e.g., in the
above example for 
$\ra = 8.0$, $\nu_r = 223\,$Hz = $\nuk/6.3$),
a particle performs several revolutions (see the plot of $\cos\phi$ in Fig.~1) 
while lingering close to the periastron, and then moves closer to the
apastron where both the rate of revolution and --- in 
particular --- the effects of the 
star's spin and quadrupole deformation (oblateness) on the orbital motion 
are much weaker.   However, when averaged over several orbital periods,  
the nodal precession rate is constant, and this {\it averaged}
value is what we denote by $\nunp$.

Equations~(\ref{ham.eq}) can be solved by direct integration over
a large number of (proper time) periods $\tau_r$ of radial motion to obtain
the nodal drift with sufficient accuracy, as was done for Fig.~2.
  More economically,
the same result can be obtained by Floquet's method (see, e.g.,
Arnol'd 1992 or Hochstadt 1975), as follows:  We define the matrix
\begin{eqnarray}
\label{Psi}
{\bf \Psi} \equiv \left( 
\begin{array}{cc}
\psi_1 & \psi_2  \\
p_1    & p_2  
\end{array} \right),
\end{eqnarray}
whose components are  
the values of the latitude $\psi$ and 
the associated momenta $p_{\psi}$ at the end of a period of radial motion,
$\psi_1 \equiv \psi\left(\tau_r\right)$, 
$p_1 \equiv p_{\psi}\left(\tau_r\right)$ for initial conditions $\psi(0) = 1$,
$p_{\psi}(0) = 0$, and $\psi_2 \equiv \psi\left(\tau_r\right)$,  $p_2
\equiv p_{\psi}\left(\tau_r\right)$ for initial conditions $\psi(0) = 0$,
$p_{\psi}(0) = 1$.  Since the product of the eigenvalues of the matrix ${\bf \Psi}$,
$\mu_1 \mu_2 = \exp\left(\int_0^{\tau_r} |{\bf A}| d\tau\right) =1$, 
we obtain
\begin{equation}
\label{mu12}
\mu_1 = \mu_2^{\ast} = e^{i\nu_{\psi}/\nu_r}, \hspace{0.4cm}
         \cos(\nu_{\psi}/\nu_r) = \frac{\psi_1 + p_2}{2},
\end{equation}
where one must include (by an appropriate addition 
of $n2\pi$) the {\it total} phase
change in $\psi$
accumulated over $\tau_r$
when calculating $\nu_{\psi}/\nu_r$ from equation~(\ref{mu12}).   The
nodal precession frequency $\nupa$ is then simply
\begin{equation}
\label{nuNP}
\nunp = \nuk - \nu_{\psi}.
\end{equation}

The constants of motion $L$ and $E$ for almost equatorial
{\it circular} orbits can be found at all $\br > \br_{\rm isco}$ 
from the equations $F(\br) = F^{\prime}(\br) =0$.  It is then
straightforward to obtain the Kepler frequency
[see equations~(\ref{dtdphi})]
\begin{equation}
\label{nukC.NS}
2\pi\nuk = \frac{L}{(E -\omega L)\br^2}\, e^{2\rho}\, +\, \omega,
\end{equation}
and the radial epicyclic frequency
\begin{equation}
\label{nurC.NS}
2\pi\nu_r = \frac{\left[- F^{\prime\prime}(\br)/2\right]^{1/2}}{E -\omega L} 
         \,  e^{\rho +\gamma}.
\end{equation}
Using equations~(\ref{I}) and (\ref{ham.eq}) we 
find the meridional frequency
\begin{equation}
\label{nupsiC.NS}
2\pi\nu_{\psi} = \frac{\Omega(\br)}{(E -\omega L)\, \br}\, e^{\gamma +\rho - \alpha},
\end{equation} 
and thus the nodal precession frequency
$\nunp = \nuk - \nu_{\psi}$.

\subsection{Kerr black holes}

Expressed in the Boyer-Lindquist coordinates,
the motion of a massive particle near the equatorial plane
of a Kerr black hole
is governed 
by the following equations (see, e.g., Chandrasekhar 1983) 
\begin{eqnarray}
\label{Kerr.dyn}
\har^4 \dot{\har}^2 &=& {\cal R},
            \nonumber  \\
\har^4 \dot{\psi}^2 &=& \Theta,
            \nonumber  \\
\har^2 \dot{\phi} &=& \frac{1}{\Delta}\left[2jE\har + L\left(\har^2 - 2\har\right)
             \right],
            \nonumber  \\
\har^2 \dot{t} &=& \frac{1}{\Delta}\left[E\Sigma^2  -2jL\har\right],
\end{eqnarray}
where
\begin{equation}
\label{del.dyn}
\Delta \equiv \har^2 - 2\har + j^2, \hspace{1.0cm} 
    \Sigma^2 \equiv \left(\har^2 + j^2\right)^2 - j^2 \Delta
\end{equation}
and
\begin{eqnarray}
\label{RTh.def}
{\cal R}(\har) &=& -\left(1-E^2\right) \har^4 + 2\har^3 - \left[j^2 \left(1 -E^2\right) + L^2
            \right] \har^2 
              \nonumber \\
         & & \hspace{2cm} + 2(L - jE)^2 \har,
         \nonumber \\  \nonumber \\
\Theta(\psi) &=&  \left[L^2 + j^2\left( 1- E^2\right)\right] \left(\psi_0^2 - \psi^2\right).
\end{eqnarray}
In the above equations we have set $\psi =0$ everywhere except
for the terms of order $\psi^2$ in the equation for the $\psi$-motion.
The amplitude $\psi_0$ of the angular displacement $\psi$ from the
equatorial plane is related to the Carter constant.

For bound orbits ($E < 1$), the equation
 ${\cal R}(\har) = 0$ has three positive roots $\hra > \hrp > \har_3$,   
and the first of equations (\ref{Kerr.dyn}) takes on the form
\begin{equation}
\label{rzeros}
\har^3 \dot{\har}^2 = - \left(1 - E^2\right) (\har -\hra) (\har -\hrp)(\har -\har_3). 
\end{equation}

Taking $\hra$ and $\hrp$ as independent variables
and defining $A \equiv \hra + \hrp $ and $B\equiv \hra \hrp$, we obtain
from the coefficients
of equation~(\ref{rzeros}) the following relations
\begin{eqnarray}
\label{coeffs}
\frac{2}{1 - E^2} &=& A + \har_3,
   \nonumber \\
j^2 + \frac{L^2}{1 - E^2} &=& B + A\har_3,
   \nonumber \\
2 \frac{(L - jE)^2}{1-E^2} &=& B\har_3.
\end{eqnarray}
Eliminating $E$ and $L$ from equations (\ref{coeffs})
we obtain a second-order equation for 
$\har_3$ 
\begin{equation}
\label{eq.r3}
D_2 \har_3^2 + D_1 \har_3 + D_0 = 0,
\end{equation}
where
\begin{eqnarray}
\label{coeffs.r3}
  D_0 &=& \left(2 B - j^2 A\right)^2
      \nonumber \\
  D_1 &=& 2 \left(2 B - j^2 A\right) \left( 2A - B - j^2\right) - 4 j^2 (A - 2) B
      \nonumber \\
  D_2 &=& \left( 2A - B - j^2\right)^2 - 4 j^2 B.
\end{eqnarray}
Once $\har_3$ is found for given $\hrp$ and $\hra$, the angular
momentum and energy are obtained straightforwardly
\begin{eqnarray}
\label{EL.Kerr}
L^2 &=& \frac{2}{\hra + \hrp + \har_3} \left(\hra\hrp + \hrp\har_3 + \har_3\hra
		- j^2\right),
       \nonumber  \\
E^2 &=& 1 - \frac{2}{\hra + \hrp + \har_3}.
\end{eqnarray}

Similarly to the neutron stars discussed above,
for $\hrp > \har_{\rm mb}$, where 
\begin{equation}
\label{rmb}
\har_{\rm mb} = 2 - j + 2\sqrt{1 -j}
\end{equation}
is the Boyer-Lindquist periastron
 of the marginally bound ($E=1$) prograde orbit,
eccentric orbits exist
only if 
$\hra > \har_{{\rm a}, {\rm min}}$  where $\har_{{\rm a}, {\rm min}}(\hrp) > \har_{\rm isco}$
and 
\begin{eqnarray}
\label{risco.Kerr}
 \har_{\rm isco} &=& 3 + Z_2 - \left[ (3 - Z_1) (3 + Z_1 + 2Z_2)\right]^{1/2},
        \nonumber \\
 & & \hspace{-0.3cm} Z_1 \equiv 1 + \left(1 - j^2\right)^{1/3} \left[ (1+j)^{1/3} +
        (1-j)^{1/3}\right],
      \nonumber \\
 & & \hspace{-0.3cm} Z_2 \equiv \left( 3j^2 + Z_1^2\right)^{1/2},
\end{eqnarray}
is the radius of the innermost  stable circular orbit.
Again,  
$\har_{{\rm a}, {\rm min}} = \hrp$ if $\hrp > \har_{\rm isco}$.

The two positive roots (if they exist, i.e., if there are
bound orbits with the given apastron $\hra$ and periastron $\hrp$) 
of equation~(\ref{eq.r3}) correspond to  prograde and retrograde 
orbits.  Restricting ourselves for the moment to weakly bound
orbits at large radii
\begin{equation}
\label{wb}
\hra, \hrp \gg 1;  \hspace{1cm}
1 + E \approx 2,    \hspace{1cm}  U \equiv 1 - E \ll 1,   
\end{equation}
around almost spherical, $j \ll 1$, black holes, we obtain from
equations~(\ref{coeffs.r3}) 
\begin{equation}
\label{r3.slow}
\har_3 \simeq 2\frac{B - j (A - 1) L}{B - 2A + 2j L}.
\end{equation}
We see that $\har_3 (L>0) < \har_3 (L<0)$ and that the  
prograde orbits are more tightly bound than the
retrograde ones, $U (L>0) > U (L<0)$.   The prograde
orbit thus corresponds to the smaller  of the
roots of equation~(\ref{eq.r3}).

The period of the radial motion is
\begin{eqnarray}
\label{Trad.Kerr}
\frac{1}{\nu_r} &=& 2 \int_{\hrp} ^{\hra} \frac{\dot{t}}{\dot{\har}}d\har
          \nonumber \\
  &=& 2\int_0^\pi \frac{E\Sigma^2 -2jL\har}
            {\Delta\sqrt{\har (1 -E^2 )(\har - \har_3)}} d\chi,
\end{eqnarray}
where we have introduced a convenient integration parameter $\chi$
such that
\begin{equation}
\label{chi.def}
 \har = \hat{l} ( 1 + \epsilon \cos\chi), \hspace{0.7cm}  
              \hat{l}  \equiv\frac{\hra +\hrp}{2},
            \hspace{0.7cm}  \epsilon \equiv\frac{\hra -\hrp}{\hra +\hrp}.
\end{equation}
Using the general relation~(\ref{nuk.NS}) to find
the Kepler frequency, we integrate the third of
equations~(\ref{Kerr.dyn})  and thus obtain 
\begin{eqnarray} 
\label{dphi.Kerr}
\Delta\phi &=& \int_{\hrp} ^{\hra} \frac{\dot{\phi}}{\dot{\har}}d\har
          \nonumber \\
  &=& \int_0^\pi \frac{2jE\har + L\left(\har^2 -2\har\right)}
            {\Delta\sqrt{\har (1 -E^2 )(\har - \har_3)}} d\chi.
\end{eqnarray}
Finally, the second of
equations~(\ref{Kerr.dyn}) and the second of equations~(\ref{RTh.def})
give the motion in the $\psi$-direction
\begin{equation}
\label{psi.Kerr}
\psi = \psi_0 \cos\left( \sqrt{L^2 + j^2\left( 1- E^2\right)} \int 
            \frac{1}{\dot{\har}}\frac{1}{\har^2} d\har\right).
\end{equation}
The failure of a particle to return to the initial $\psi$ 
after a full period of azimuthal motion will lead to a 
secular precession of the orbital plane (i.e., the nodal precession)
at frequency (\ref{nuNP})
where 
\label{nupsi.K}
\begin{eqnarray}
\label{Tpsi.Kerr}
\nu_{\psi} &=& \nu_r 
               \frac{1}{\pi} \sqrt{L^2 + j^2\left( 1- E^2\right)} \int_{\hrp} ^{\hra}
            \frac{1}{\dot{\har}}\frac{1}{\har^2} d\har
             \nonumber \\
         &=& \nu_r \frac{1}{\pi} \sqrt{L^2 + j^2\left( 1- E^2\right)}
               \nonumber \\
           & & \hspace{0.3cm} \times  \int_0^\pi \frac{1}
              {\sqrt{\har^5 (1 -E^2 )(\har - \har_3)}} d\chi. 
\end{eqnarray}

For a circular orbit at Boyer-Lindquist radius $\har$
near the equatorial plane,
the Kepler frequency for prograde orbits 
is (see, e.g., Chandrasekhar 1983)
\begin{equation}
\label{Kepler}
2\pi \nuk =  \frac{1}{\har^{3/2}\left(1 + j \har^{-3/2}\right)}
\end{equation}
(to obtain expressions for retrograde orbits one
should set  $j \rightarrow -j$).
At all $\har > \har_{\rm isco}$
a slight perturbation of the orbital motion in the radial
direction will lead to radial oscillations at the
radial epicyclic frequency
\begin{equation}
\label{radial.epi}
\nu_r = |\nuk| \left(1 - 6 \har^{-1} + 8 j \har^{-3/2}
- 3j^2 \har^{-2}\right)^{1/2}.
\end{equation}

On the other hand, a particle in a circular orbit tilted slightly
with respect to the equatorial plane will undergo meridional
oscillations at frequency
\begin{equation}
\label{nu.psi}
\nu_{\psi} = |\nuk | \left(1 - 4j \har^{-3/2} + 3j^2 \har^{-2}\right)^{1/2}
\end{equation}
leading to the nodal precession at 
frequency (\ref{nuNP}).

At large radii ($r \agt 10$), the asymptotic expressions
$2\pi\nupa \approx 3 \har^{-5/2}$ and $2\pi\nunp \approx 2j \har^{-3}$
lead to the simple scaling
\begin{equation}
\label{pa.lt}
\nunp \approx j \pi^{1/5} \left(\frac{2}{3}\right)^{6/5} \nupa^{6/5}.
\end{equation}
At smaller radii, the slope $d\ln\nunp/d\ln\nupa$ steadily decreases with
increasing frequencies and drops below unity near the innermost
stable  circular
orbit.

\subsection{Spinning stars in post-Newtonian approximation}

At a sufficiently large distance from a spinning star of 
angular momentum $j GM^2/c$ and  quadrupole moment
$q G^2 M^3/c^4$ ($q < 0$ for oblate bodies) we can use
post-Newtonian (PN) expansions (see, e.g., Weinberg 1972) to obtain approximate expressions
for the periastron advance and nodal precession
frequencies.   Thus, for a nearly equatorial
orbit of semilatus rectum $l \equiv 2\ra\rp/ (\ra - \rp)$, eccentricity 
$\epsilon \equiv (\ra - \rp)/(\ra + \rp)$, and the Newtonian and post-Newtonian
orbital (Kepler) frequencies
\begin{eqnarray}
\label{nuk.N}
\nukN &=& \frac{1}{2\pi} \left(\frac{1 - \epsilon^2}{l}\right)^{3/2},
   \nonumber \\
\nuk &=& \nukN \left[ 1 - \left(1-3 \epsilon^2\right)\frac{1}{l} + 
            {\cal O}\left(\frac{1}{l^{3/2}}\right) \right],
\end{eqnarray}
the
rate of periastron advance is
\begin{eqnarray}
\label{pa.PN}
\nupa &=& \nuk \bigg[ 3\frac{1}{l} + \frac{9}{2}
            \left(1 + \frac{\epsilon}{6}\right) \frac{1}{l^2} 
                  \nonumber \\
         & & \hspace{1cm}     - 4j\frac{1}{l^{3/2}} - \frac{3q}{2}\frac{1}{l^2} +  
                {\cal O}\left(\frac{1}{l^{5/2}}\right) \bigg],
\end{eqnarray}
where ${\cal O}\left(\frac{1}{l^{5/2}}\right)$ indicates
terms of asymptotic order at least as high as $l^{-5/2}$.
In equation~(\ref{pa.PN}),
 the first two terms are the 1$^{\rm st}$ and 2$^{\rm nd}$
PN corrections present even for a spherical star, the
third term is due to the gravitomagnetic force, while
the fourth term is the (Newtonian) effect of the star's
rotation-induced deformation. 

In the lowest-order PN approximation, the precession of
the plane of an orbit around the star's spin axis is
interpreted as being caused by the gravitomagnetic force.
Together with the purely Newtonian quadrupole precession,
the nodal precession rate is 
\begin{equation}
\label{np.PN}
\nunp = \nuk \left[ 2j\frac{1}{l^{3/2}}  + \frac{3q}{2} \frac{1}{l^2}
                + {\cal O}\left(\frac{1}{l^{5/2}}\right) \right].
\end{equation}

While for a Kerr black hole $q = -j^2$ [cf. equations (\ref{Kepler}) and
(\ref{nu.psi})], the ratio $-q/j^2$ is considerably
larger for neutron stars (Laarakkers \& Poisson 1998); it  ranges from $\sim 2$
to more than 10, depending on the mass and the equation of state.

\begin{table}
  \begin{center}
    \begin{tabular}{cccccccccc}\hline \hline
      & & \hspace{0.5cm} {\rm C} \hspace{0.6cm} 
        & \hspace{0.5cm} {\rm UU} \hspace{0.6cm} 
        & \hspace{0.5cm} {\rm L} \hspace{0.6cm} \\
    \hline 
      & &  \hspace{-0.1cm} $\Re$ \hspace{0.2cm}  $\risco$ \hspace{0.0cm} 
        &  \hspace{-0.1cm} $\Re$ \hspace{0.2cm}  $\risco$ \hspace{0.0cm}
        &  \hspace{-0.1cm} $\Re$ \hspace{0.2cm}  $\risco$ \hspace{0.0cm} \\
      & &  \hspace{-0.1cm} (km) \hspace{0.0cm}  (km) \hspace{0.0cm}
        &  \hspace{-0.1cm} (km) \hspace{0.0cm}  (km) \hspace{0.0cm}
        &  \hspace{-0.1cm} (km) \hspace{0.0cm}  (km) \hspace{0.0cm} \\
    \hline 

   & $\hspace{-0.5cm} j=0.1$  & \hspace{-0.4cm} 4.08 \hspace{0.2cm} 5.69  
             & \hspace{-0.4cm} 4.13 \hspace{0.2cm} 5.69  
             & \hspace{-0.4cm} 5.69 \hspace{0.2cm} 5.72  \\   
	\vspace{0.1cm}
   & $\hspace{-0.5cm} $     & \hspace{-0.4cm} (10.8) \hspace{0.0cm} (15.1)
             & \hspace{-0.4cm} (11.0) \hspace{0.0cm} (15.1)  
             & \hspace{-0.4cm} (15.1) \hspace{0.0cm} (15.2)  \\   
   & $\hspace{-0.5cm} j=0.2$  & \hspace{-0.4cm} 4.17 \hspace{0.2cm} 5.41  
             & \hspace{-0.4cm} 4.19 \hspace{0.2cm} 5.42  
             & \hspace{-0.4cm} 5.76 \hspace{0.2cm} - - -   \\
   & $\hspace{-0.5cm} $     & \hspace{-0.4cm} (11.1) \hspace{0.0cm} (14.4)
             & \hspace{-0.4cm} (11.1) \hspace{0.0cm} (14.4)
             & \hspace{-0.5cm} (15.3) \hspace{0.10cm} - - -  \\

         \hline
     & &  \hspace{0.2cm} $\nus$ \hspace{0.4cm}  $q$ \hspace{0.3cm}
        &  \hspace{0.2cm} $\nus$ \hspace{0.4cm}  $q$ \hspace{0.3cm}
        &  \hspace{0.2cm} $\nus$ \hspace{0.4cm}  $q$ \hspace{0.3cm} \\
      \hline
   & $\hspace{-0.5cm} j=0.1$  & 271 \hspace{0.1cm} -0.024
             & 249 \hspace{0.1cm} -0.026
             & 148 \hspace{0.1cm} -0.056  \\
   & $\hspace{-0.5cm} j=0.2$  & 525 \hspace{0.1cm} -0.090
             & 490 \hspace{0.1cm} -0.099
             & 292 \hspace{0.1cm} -0.209   \\
         \hline

    \end{tabular}
  \end{center}
\caption{
   Stellar equatorial (circumference) 
radii $\Re$, ISCO radii,  spin frequencies $\nus$ [Hz] and quadrupole moments $q$ 
in units $G^2M^3/c^4$ for 
NS models of mass $M= 1.8$ and indicated angular momenta $j$.   The radii are
given both in units $GM/c^2$ and in kilometres.
}
\end{table}

\section{Bound orbits around neutron stars: comparison with Kerr black hole orbits}

Equations of state of different stiffness give rise to a wide
range of magnitude of the model neutron stars'
rotation-induced deformation, and,
consequently, significant differences in the spacetime
geometries around them.  It is our main purpose
in this Section to describe and quantify how
the differing geometries  imprint themselves
on the potentially measurable frequencies of
bound orbits around the stars, and to compare these
frequences with those of orbits around spinning (Kerr)
black holes.

For neutron star models of given mass $M$
and angular momentum $j GM^2/c$, the very stiff
equations of state
(e.g., EOS L as
opposed to the `softer' UU or yet `softer' C)
yield more extended neutron star models
that undergo greater
centrifugal deformation.  This leads to
higher
mass quadrupole moments $q G^2M^3/c^4$ while
requiring lower spin frequencies $\nus$,
as can be seen in Table~1.

\begin{figure}
\label{Kerr.circ}
\centering
\hbox{\epsfxsize= 10.5 cm \epsfbox[20 150 370 710]{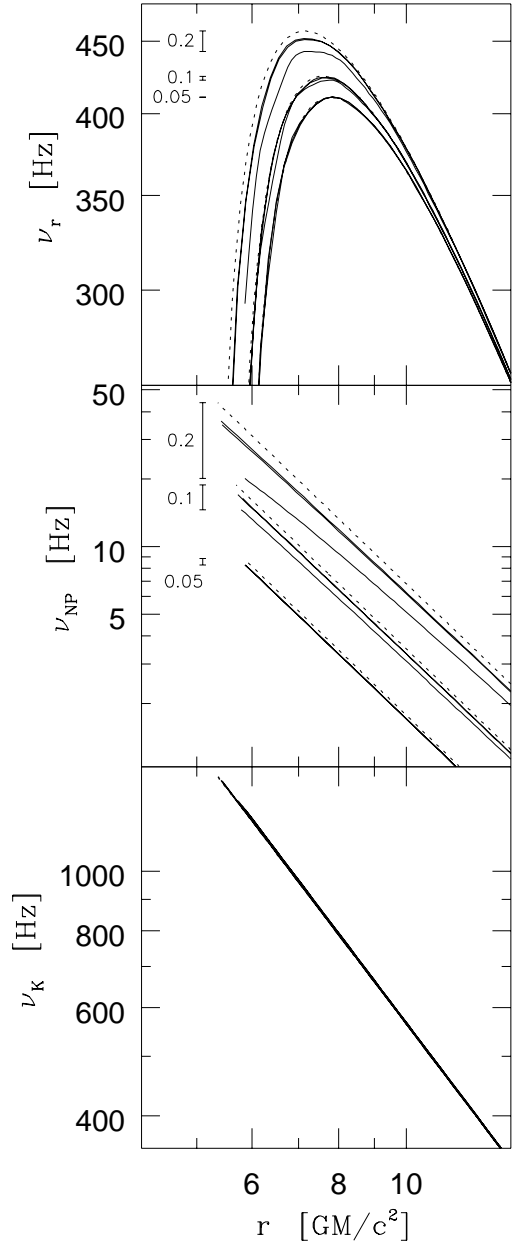}}
\caption{The radial, nodal precession, and Kepler frequencies as
functions of the circumference radius for {\it circular} orbits
around  a $M=1.8$ neutron stars (solid lines;
EOS C, UU,, and L) and Kerr black holes (dotted lines).  The
values of the angular momentum parameter  $j$
are indicated by the labels of vertical bars.
}
\end{figure}

As Laarakkers \& Poisson (1998, LP) have shown, 
the magnitude of $q$ grows nearly quadratically, 
$q \simeq - a({\rm EOS}, M) j^2$, over a broad range
of $j$.   While the constant $a$ is much smaller for
EOS C ($a\simeq 2.4$ for $j \alt 0.2$; LP compute $a\simeq 2.7$
for a much greater range of $j$) or EOS UU ($a\simeq 2.6$) than
for the stiff EOS L ($a\simeq 5.2$; LP find $a\simeq 4.9$), it still exceeds
by a large amount the (exact) value $a = 1$ for the Kerr black
hole.

\begin{figure}
\label{nuPA.cir}
\centering
\hbox{\epsfxsize= 10.5 cm \epsfbox[20 490 370 710]{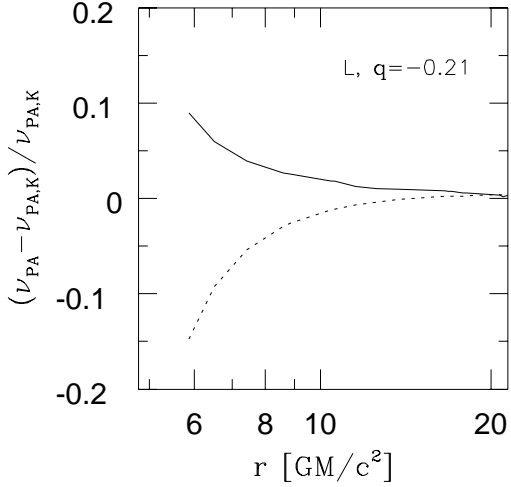}}
\caption{
Relative deviation, $(\nupa - \nu_{_{\rm PA,K}})/\nu_{_{\rm PA,K}}$,
with respect to the Kerr values (for the same $M$ and $j$)
of the periastron advance frequencies (solid line) and the post-Newtonian
values (dotted line; see equation \ref{pa.PN}) for circular orbits
around an $M=1.8$, $j=0.2$, EOS L ($q=-0.21$)  neutron star.
}
\end{figure}

\begin{figure}
\label{nuNP.cir}
\centering
\hbox{\epsfxsize= 10.5 cm \epsfbox[20 490 370 710]{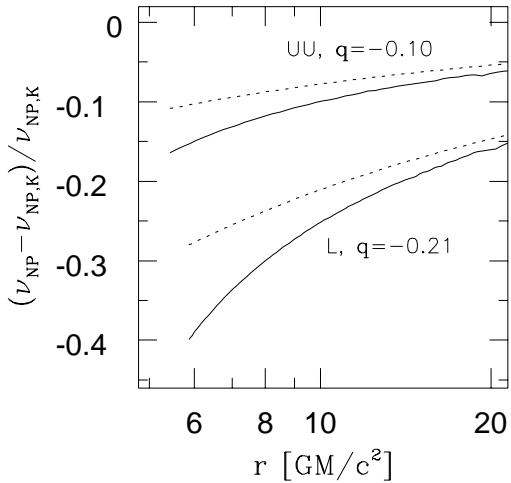}}
\caption{Relative deviations, $(\nunp - \nu_{_{\rm NP,K}})/\nu_{_{\rm NP,K}}$,
with respect to the Kerr values (for the same $M$ and $j$)
of the nodal precession frequencies (solid lines) and the post-Newtonian
values (dotted lines) 
for circular orbits
around $M=1.8$, $j=0.2$ neutron stars.
The EOS used are UU (yielding $q=-0.10$) for the upper 
solid line, and $L$
($q=-0.21$) for the lower solid line.
The post-Newtonian values are obtained from 
equation (\ref{np.PN}) using the values
of $M$, $j$ and $q$ of the corresponding neutron star models.
}
\end{figure}

Based on the low-order PN expressions~(\ref{pa.PN}) and
(\ref{np.PN}), one would expect a general trend toward higher
$\nupa$ (i.e., lower $\nu_r$) and lower $\nunp$ at given $r$ (all
radii used in this section are the coordinate-independent
 {\it circumferential} radii) as
$-q$ is increased.   This expectation is indeed borne out by
our computation of circular orbital frequencies as can be seen,
e.g., in Fig.~3 for a $M=1.8$ family
of NS models and black holes.   Of course,
a PN expansion cannot account for the strong-field
relativistic effect of the vanishing of the radial
epicyclic frequency at the radius $\risco$
of  the innermost stable circular orbit; the
large magnitude
of $q$ for a stiff EOS simply causes a faster drop of $\nu_r$ as
$r$ is reduced past the point of maximum $\nu_r$.  Quite generally, $\risco$
moves inward for higher $j$, an effect familiar from
the case of the Kerr spacetime~(\ref{risco.Kerr}).

The vertical bars in Fig.~3 mark the ranges of the maximal values
of $\nu_r$ (the top panel) and $\nunp$ (the middle panel) for
the indicated values of $j$.  For each $j$ we plot
the frequencies for the three EOS, C, UU and L (all solid lines),
and the Kerr black hole (the dotted line)
of the same $M$ and $j$.   In all cases
the curves
for EOS C and UU almost coincide.
Whereas the drop in $\nu_r$ at $r < 10$ due to the star's spin-induced
deformation
is relatively modest [$\,|\Delta\nu_r|/\nu_r\alt 15\%$;
notice, however the considerable numerical error in $\nu_r$ for the stiff EOS L,
due
to the necessity of computing
the second derivative (\ref{nurC.NS}) from the
tabulated metric],
the reduction in $\nunp(r)$ relative to the Kerr values
is more pronounced, amounting at $j=0.2$ to
up to $\sim 20\%$  for EOS UU or even $\sim 40\%$ for EOS L.
For lower $j$ ($j < 0.1$), the difference between
the frequences for different EOS is less pronounced, and
it virtually disappears at $j\sim  0.05$.

\begin{figure}
\label{Kerr.exc}
\centering
\hbox{\epsfxsize= 10.5 cm \epsfbox[20 150 370 710]{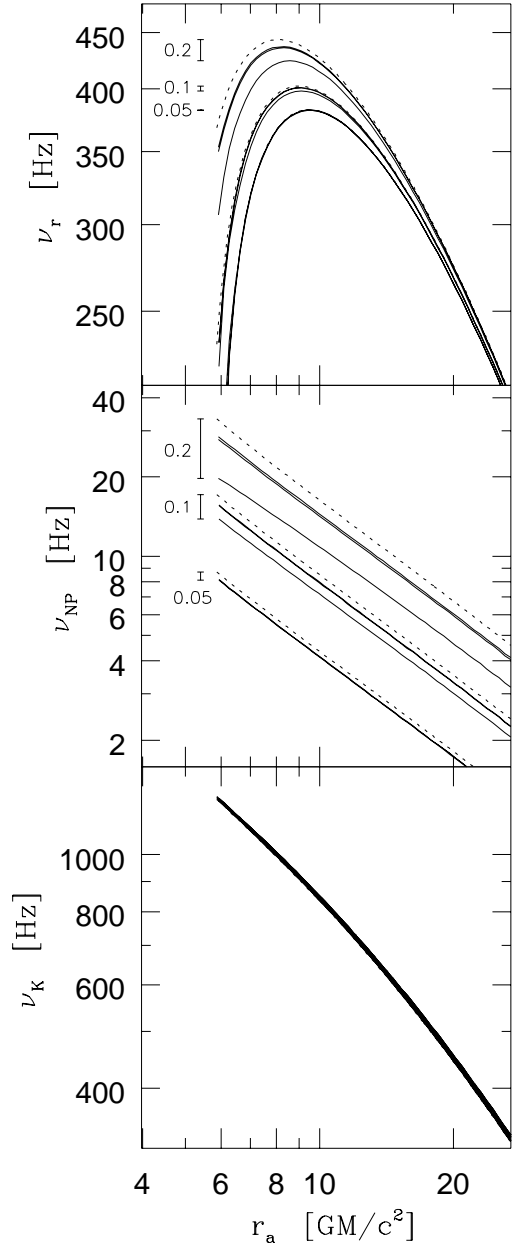}}
\caption{The radial, nodal precession and Kepler frequencies as
functions of the apastron radius for eccentric orbits of common periastron
$r_p = 5.88$
around  $M=1.8$ neutron stars (solid lines;
EOS C, UU and L) and Kerr black holes (dotted lines).  The
values of the angular momentum parameter  $j$
are indicated by the labels of the vertical bars.
}
\end{figure}

\begin{figure}
\label{ratfr}
\centering
\hbox{\epsfxsize= 10.5 cm \epsfbox[10 430 340 710]{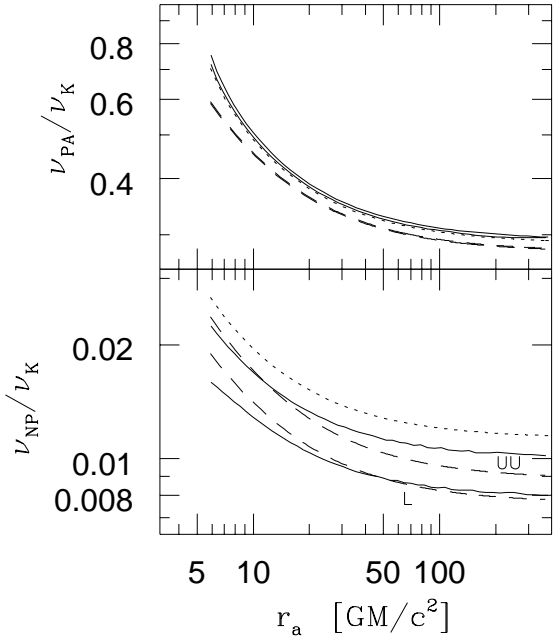}}
\caption{Ratios of orbital frequencies for $M= 1.8$, $j=0.2$ and
$\rp =5.88$
for EOS L and UU (solid lines) and for the Kerr black hole (dotted lines)
of the same $M$ and $j$.  PN values for $q=-0.21$ (EOS L)   
and $q=-0.10$ (EOS UU) are given by dashed lines.
}
\end{figure}


\begin{figure}
\label{nuPA1.exc}
\centering
\hbox{\epsfxsize= 6.0 cm \epsfbox[137 475 332 680]{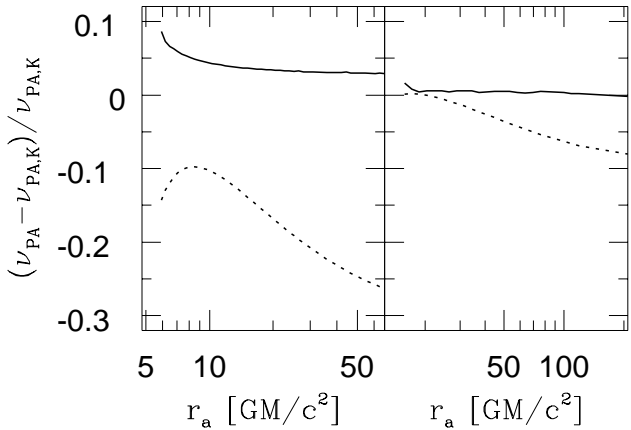}}
\caption{
Relative deviations, $(\nupa - \nu_{_{\rm PA,K}})/\nu_{_{\rm PA,K}}$,
with respect to the Kerr values (for the same $M$ and $j$)
of the periastron advance frequencies (solid lines) and the post-Newtonian
values (dotted lines; see equation \ref{pa.PN}) for eccentric orbits
around an $M=1.8$, $j=0.2$, EOS L ($q=-0.21$) neutron star.
{\it Left panel:} eccentric orbits
of common periastron $\rp= 5.88$.  {\it Right panel:} eccentric orbits
of common periastron $\rp= 15.0$.  
}
\end{figure}




Of the three EOS, the stiff EOS L makes the star at $j=0.2$ so
extended at the equator, $R_{\rm e} = 5.8 GM/c^2 = 15.4\,$km,
that there is no innermost stable circular orbit (see Table~1);
the corresponding curve $\nu_r(r)$ thus never reaches the
terminal point $\nu_r=0$, in contrast with all the
other cases shown in Fig.~3.

The simple low-order PN expressions (\ref{pa.PN}) for $\nupa$ and
(\ref{np.PN}) for $\nunp$
approximate rather accurately the  actual frequencies of circular orbits
of radii $r\agt 10$, as can be seen from Figs.~4 and 5.  As pointed
out above, close to $\risco$  (i.e., for $r\alt 8$), the
growth of the post-Newtonian $\nupa$ with decreasing $r$ cannot model
accurately the much faster growth of the relativistic $\nupa$ as it
approaches $\nuk$.   In this range, the Kerr values for the same $M$
and $j$ furnish a considerably better approximation, as is evident in Fig.~4.

For the nodal precession, on the other hand,
the large quadrupole term
makes at all $r$ the low-order expansion~(\ref{np.PN}) a much better
approximation than the Kerr $\nunp(r)$ (see Fig.~5).
Notice, however, that for stiff EOS and
for $r \alt 8$ (the lower curves of Fig.~5)
the expansion~(\ref{np.PN}) significantly overestimates the actual
$\nunp$ which is, e.g., at the surface of the star,
less than $60\%$ of the corresponding Kerr value.

\begin{figure}
\label{nuNP.exc1.2}
\centering
\hbox{\epsfxsize= 6.0 cm \epsfbox[137 475 332 680]{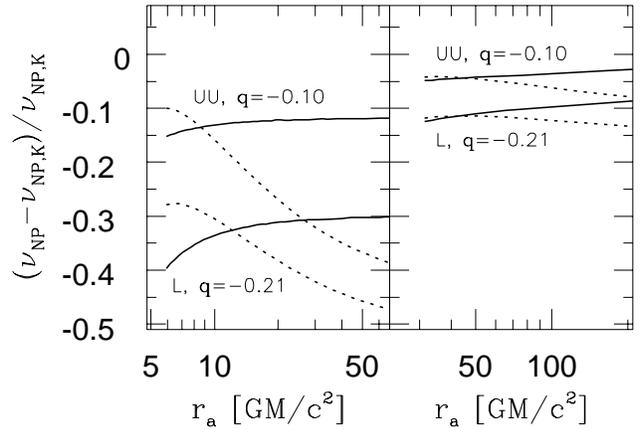}}
\caption{
Relative deviations, $(\nunp - \nu_{_{\rm NP,K}})/\nu_{_{\rm NP,K}}$,
with respect to the Kerr values (for the same $M$ and $j$)
of the nodal precession frequencies (solid lines) and the post-Newtonian
values (dotted lines; see equation \ref{np.PN}) for eccentric orbits
around $M=1.8$, $j=0.2$ neutron stars.
The EOS used are UU (yielding $q=-0.10$) for the upper solid and dotted lines,
and $L$
($q=-0.21$) for the lower lines. {\it Left panel:} eccentric orbits
of common periastron $\rp= 5.88$.  {\it Right panel:} eccentric orbits
of common periastron $\rp= 30.0$.
}
\end{figure}

\begin{figure}
\label{Kerr.excE}
\centering
\hbox{\epsfxsize= 10.5 cm \epsfbox[20 150 370 710]{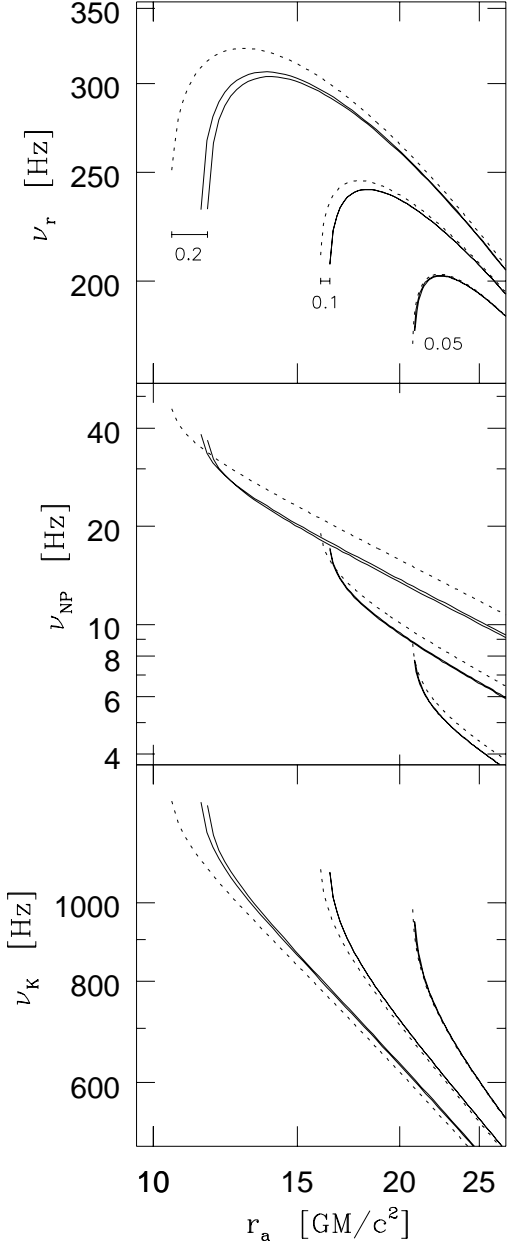}}
\caption{The radial, nodal precession and Kepler frequencies as
functions of apastron radii of highly eccentric orbits of common periastron
$r_p = 4.30$
around  $M=1.8$ neutron stars (solid lines;
EOS C and UU) and Kerr black holes (dotted lines).  The
values of the angular momentum parameter  $j$
are indicated in the top panel.
}
\end{figure}

\begin{figure}
\label{UU.demo}
\centering
\hbox{\epsfxsize= 10.5 cm \epsfbox[40 380 390 710]{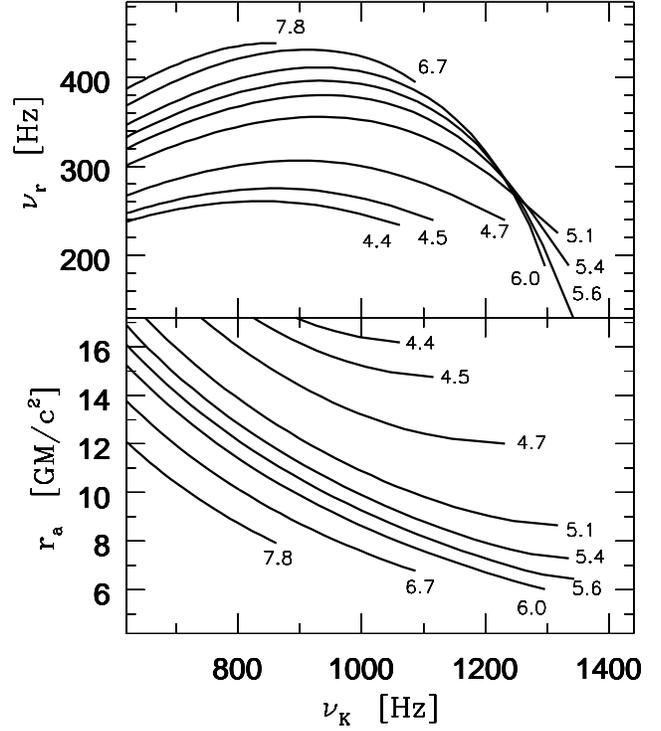}}
\caption{EOS UU:  frequencies and apastra as
functions of $\nuk$ for families of eccentric
orbits around a NS of $M= 1.7$ and $\nus = 100\,$Hz
($Re = 4.39 = 11.0\,$km, $\risco = 5.87 = 14.7\,$km).  
The curves are
labeled by the values of periastron in units $GM/c^2$.
}
\end{figure}

\begin{figure}
\label{L.demo}
\centering
\hbox{\epsfxsize= 10.5 cm \epsfbox[40 380 390 710]{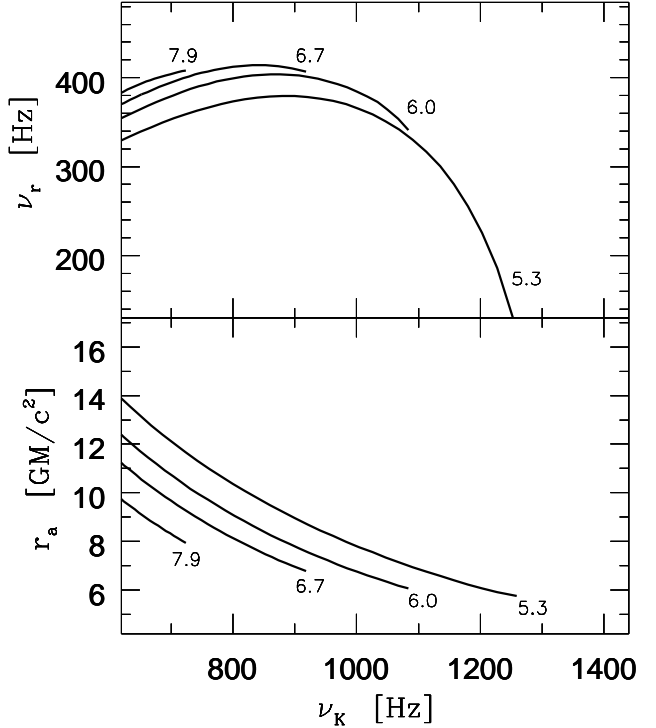}}
\caption{EOS L:  frequencies and apastra for families of eccentric
orbits around a NS of $M= 2$ and $\nus = 363\,$Hz  ($\Re = 5.23 = 15.4\,$km,
$\risco = 5.44 = 16.1\,$km). 
}
\end{figure}

Turning now to eccentric orbits, one can distinguish between two broad classes,
characterised by periastra $\rp$ that are either larger or
smaller  than $\risco$.   
Within the first class, $\rp > \risco$ (so that $r_{\rm a,min} = \rp$),
the plots of the orbital frequences  as functions of $\ra$
for a family
of orbits of common periastron $\rp$ resemble (an example
is shown in Fig.~6)
the frequency vs. radius plots for circular orbits shown above in Fig.~3.
The obvious difference is that the magnitudes of the negative slopes $d\nu/d\ra$ are 
smaller than those of the slopes $d\nu/dr$ for circular orbits: even
if $\ra$ is large, the
eccentric orbits have access to small radii where relativistic
effects, like frame dragging, take place.  This can be seen from the
PN expansions~(\ref{pa.PN}) and (\ref{np.PN}), in which the effect of
increased periastron radius  is `buffered' 
by $\ra$'s appearing in the expressions only through
$l$, where $l \sim 2 \rp$
for $\ra \gg \rp$ [the dependence on $\epsilon$ in (\ref{pa.PN}) is weak]. 
For large $\ra$ we therefore expect the ratios $\nupa/\nuk$ and
$\nunp/\nuk$ to asymptote to constant values, as indeed is
the case (see Fig.~7).

Although the PN expressions~(\ref{pa.PN}) and (\ref{np.PN}) reproduce
qualitatively the profiles $\nupa(\ra)$ and $\nunp(\ra)$ obtained
numerically from NS models, they tend to give a 
somewhat steeper
fall-off at large $\ra$, thus falling short of the `exact' values
by up to $\sim 30\%$.  This is evident in Fig.~7 and is displayed
in greater detail in Figs.~8 and 9.  It is a consequence of the
extreme slowness of relativistic
radial motion in the strong-filed vicinity of
$\risco$: since relativistic effects (e.g., frame dragging)
weaken rapidly
at larger radii, this gives them 
more time to build up and the integrated effects thus exceed
the PN estimates.  

As in the case of circular orbits, the periastron advance 
frequency $\nupa(\ra)$ is modeled much better for all $\ra$ by the
Kerr black hole values (for the same $M$ and $j$;
see Fig.~8) than by the PN expansion
(\ref{pa.PN}).  As can be seen in the right panel of Fig.~8,
only for $\rp > 10$ does the PN $\nupa(\ra)$ approach closely
the values obtained from the numerically computed NS spacetimes.

Regarding the nodal precession, for moderately eccentric orbits,
$\ra \alt 2\rp$, the PN expansion~(\ref{np.PN}) is superior
to the Kerr values at modeling $\nunp(\ra)$ obtained from
the numerical NS spacetimes (see Fig.~9).  The better
accuracy of the PN frequences is more prominent for stiff
EOS, in a straightforward
extension of the trend we observed for circular orbits.
Again, for larger eccentricities, the numerical spacetimes' 
$\nunp(\ra)$ inevitably split off from the PN curves and
shift toward the Kerr values.  As we increase
$\rp$,  the range 
of eccentricity $\epsilon$ over which
the PN expansion is relatively
accurate increases; compare the two panels of
Fig.~9.

For the second class of eccentric orbits with $\rp < \risco$,
we observe (see Fig.~10 for EOS C and UU) 
that at constant $\rp$  the minimum
apastron radius, $r_{\rm a, min} (\rp)$,
shrinks rapidly with increasing $j$, following the related
reduction of $\risco$  (for both EOS, $\risco = 5.84,$ $5.69$ and $5.42$
for $j = 0.05$, $0.1$ and $0.2$, respectively).  Compared
with the Kerr black hole, the main effect of the larger
quadrupole moment of the NS is a larger $r_{\rm a, min} (\rp)$
at given $M$ and $j$ as well as  lower $\nu_r(\ra)$
and higher $\nupa(\ra)$.  Notice, however, that as $\ra$ approaches $r_{\rm a, min}$
from above, the functions  $\nuk(\ra)$ and $\nunp(\ra)$ bend 
upward abruptly, while $\nu_r(\ra)$ drops
until it reaches the `terminal' value $\nu_r (r_{\rm a, min})$.

The sharp change of $\nuk$ near  $r_{\rm a, min}$ leads to a
`flat' dependence $\nu_r(\nuk)$ for a family of orbits of a given
$\rp < \risco$;
see Fig.~11 for a relatively
slow spinning, $\nus=100\,$Hz, EOS UU neutron star. 
As $\rp$ is increased, the functions $\nu_r(\nuk)$ at first
grow steeper, with increasingly negative slopes at large $\nuk$
(small $\ra$), until at $\rp = \risco$, $\nu_r$ drops to
zero for $\nuk = \nuisco$.
A further growth of $\rp$ again brings about flatter functions
$\nu_r(\nuk)$ but at considerably larger $\nu_r$.

In Fig.~12 we show the functions $\nu_r(\nuk)$ for eccentric
orbits around a $M=2$, $\nus =363\,$Hz, EOS L neutron star.  The  
stiff equation of state makes the star large ($R_{\rm e} = 5.3 =
15.6\,$km) so that very low $\rp$ orbits do not exist
and the flat, low $\nu_r$ curves $\nu_r(\nuk)$ seen in Fig.~11
are absent from Fig.~12.   (This is why the plots
of frequencies for the low $\rp =4.3$ families of orbits in Fig.~10 could
be shown only for the {\it moderately} stiff EOS C and UU.)  As a consequence,
the set of frequencies of eccentric orbits (for given $M$ and $\nus$) is a much
more limited subset of the 2-dimensional $\nuk$-$\nu_r$ plane for the stiff EOS L
than for the softer equations of state.

\section{Conclusions}

If shown to indeed be a manifestation of the orbital motion
around neutron stars, the quasi-periodic oscillations
of the low mass X-ray binaries would furnish an unprecedented probe of
the strongly curved spacetime
near the compact objects and allow inference of their masses $M$ and spin rates $\nus$.
In particular, highly eccentric orbits, which near their periastra penetrate deeper
than the innermost stable circular orbit, would put tight constraints on the
sizes of neutron stars, and thus on the equation of state.

The high accuracy (e.g., better than $\sim 1\%$ for Sco~X-1), with which
the frequencies of the kHz QPO frequencies have been measured,  makes it
important to use the values obtained by integrating test-particle motion
in fully relativistic numerical spacetimes that can be computed by spinning neutron
star codes.  For $j > 0.1$, and especially for orbits of considerable eccentricity,
the post-Newtonian (PN) expansions or approximation with Kerr
metric orbital frequencies are not sufficiently accurate.

In the case of the periastron advance frequency $\nupa$,
the Kerr results provide a
better approximation to actual motion in the neutron star spacetimes
than do the low-order PN expressions.  Although the Kerr $\nu_{_{\rm PA,K}}$ is always
lower than the value $\nupa$ for the same periastron and apastron 
(circumferential) radii in a
neutron-star spacetime, it never falls below $\nupa$ by more than 10\%, even
for the highest angular momenta
investigated here ($j\approx 0.2$) and for the stiffest EOS (L).
The general-relativistic decrease of $\nu_r$ in the vicinity of
the innermost stable orbit (ISCO) makes the PN 
approximations for $\nupa$ inaccurate even
for moderately ($\ra/\rp \alt 2$)
eccentric orbits if $\Rp < 7$ (for highly eccentric orbits, 
PN is typically a poor approximation).

The nodal precession frequency $\nunp$ of 
orbits inclined relative to the equatorial plane
of a neutron star is dominated by the prograde gravitomagnetic `torque,'
with a large retrograde (for prograde orbits) contribution from the classical
quadrupole torque.  In the cases considered in this paper, the latter
torque can amount at small orbital radii from about $10\%$ to up to
about $50\%$ of the magnitude of the gravitomagnetic torque.
The spin-induced deformation (oblateness) of the neutron stars renders, therefore,
the low-PN $\nunp$ far more adequate (for orbits of moderate eccentricity)
than the values obtained from Kerr
black holes with the same $M$ and $j$.   However, for orbits of very small
size ($\Ra \alt 8$), even the low-order PN expressions overestimate $\nunp$
by up to 30\% in the case of $j\approx 0.2$, stiff EOS L models.

Very stiff equations of state (e.g., L) have an important
effect on the high-frequency portion of the
set of all available frequencies $\nuk$, $\nu_r$
of eccentric orbits: due to the large stellar equatorial radii,
orbits with very low periastra
do not exist.  The consequent absence of low values of $\nu_r$ at $\nu_{\rm K, isco}/2 \alt
\nuk < \nu_{\rm K, isco}$, for given $M$ and $\nus$,
restricts severely the available subset of the $\nuk$-$\nu_r$ plane.  
For this reason,  
very stiff equations of state do not permit
successful modeling of the neutron-star
LMXBs' kHz QPO frequencies 
in terms of the frequencies of eccentric orbits (see Markovic \& Lamb 2000).

\subsection*{Acknowledgements}
The author is indebted to Sharon Morsink, Scott Hughes, Fred Lamb and Luca Zampieri 
for enlightening discussions of conceptual and technical issues and to Nikolaos
Stergioulas for making his code available.  
He has also benefited from conversations
with Norman Murray, Vijay Pandharipande and Geoffrey Ravenhall.
The Aspen Institute for Physics has provided
a  pleasant environment for interaction with colleagues.  This work was
supported in part by NSF grant 
AST~96-18524 and by NASA grant NAG~5-2925 at Illinois.

\end{document}